\documentclass{svproc}
\usepackage{url}

\usepackage{amsmath,amssymb,amsfonts}
\usepackage{graphicx}
\usepackage{float}
\usepackage{multirow}
\usepackage{xcolor}

\begin{document}
\mainmatter              
\title{Identification For Control Based on Neural Networks: Approximately Linearizable Models}
\titlerunning{Identification of Approximately Linearizable Models}  
%
\author{Maxime Thieffry, Alexandre Hache, Mohamed Yagoubi, Philippe Chevrel}
\authorrunning{M. Thieffry et al.} 
%
\tocauthor{~}
\institute{IMT Atlantique, LS2N, UMR 6004, F-44300, Nantes, France \hspace{7pt}
\email{firstname.lastname@imt-atlantique.fr}}

\maketitle              

\begin{abstract}
This work presents a control-oriented identification scheme for efficient control design and stability analysis of nonlinear systems. Neural networks are used to identify a discrete-time nonlinear state-space model to approximate time-domain input-output behavior of a nonlinear system. The network is constructed such that the identified model is approximately linearizable by feedback, ensuring that the control law trivially follows from the learning stage. After the identification and quasi-linearization procedures, linear control theory comes at hand to design robust controllers and study stability of the closed-loop system. The effectiveness and interest of the methodology are illustrated throughout the paper on popular benchmarks for system identification.
\end{abstract}
\section{Introduction}

\subsection{Context and Related Works}

The ongoing growth of data and computing resources prompts all scientific communities to take an interest in data-driven methods, and automatic control is no exception \cite{alamir2022learning}. However, while modern machine learning techniques can handle very large datasets, the underlying dynamics renders data generated by dynamical systems temporally correlated which makes control-oriented learning algorithms less efficient \cite{sznaier2020control}. Whereas different literature works aim at either reducing the computational complexity of
the learning problem or improving model identification performances, see e.g. \cite{schon2011system,schoukens2019nonlinear,forgione2020model}, our objective is slightly different as we seek to provide a control-oriented identification method to ease the observer and/or controller design process. Control-oriented learning takes different forms but one can find among others: the identification of linear parameter varying (LPV) approximation of nonlinear models \cite{ghosh2018optimal,verhoek2022}, the use of Koopman theory to learn a linear representation of a nonlinear system \cite{mauroy2016linear,lusch2018deep} or learning-based model predictive control \cite{hewing2020learning}. 

We consider a plant whose evolution is described by an unknown discrete-time nonlinear state-space model of the form:

\begin{equation}
    \begin{aligned}
    x(t+1)&=f(x(t), u(t))\\
    y(t)&=Cx(t)
    \end{aligned}
    \label{eq:modelNL}
\end{equation}
where $u(t) \in \mathbb{R}^{m}$ is the system input, $y(t) \in \mathbb{R}^{p}$ the output, $x(t) \in \mathbb{R}^{n}$ is the state and where the function $f : \mathbb{R}^n \times \mathbb{R}^m \rightarrow \mathbb{R}^n$ is unknown and generally nonlinear. For sake of clarity, results are presented for the linear-output case, where the matrix $C$ from equation \eqref{eq:modelNL} is unknown. Discussion at the end of the paper shows how to extend the results to models with nonlinear output terms.

A neural state-space model is a model where the function $f$  from equation \eqref{eq:modelNL} is approximated using a neural network $f_n$ with $l$ hidden layers of the form:
\begin{equation}
    \begin{aligned}
        z_0 &= z \\
        z_{i+1} &= \sigma_{i+1}(W_i z_i+b_i)~,~i = 0 , \hdots, l-1 \\
        f_n(z) &= W_l z_l+b_l
    \end{aligned}
\end{equation}
where $W_i$ and $b_i$ are weights matrices and biases vectors respectively
and $\sigma_i$ is an appropriate nonlinear activation functions, usually sigmoid or hyperbolic tangent.

The inclusion of an explicit linear part in the nonlinear model has been studied by several authors that show how it improves the identification of the nonlinear model, see e.g. \cite{sjoberg1997estimation,schoukens_initvieux} The model to be identified is then written as:
\begin{equation}
    \begin{aligned}
    x(t+1) &= Ax(t)+Bu(t)+f_{n}(x(t),u(t)) \\
    y(t) &= Cx(t) 
    \end{aligned}
    \label{eq:ssnn_schoukens}
\end{equation}
This model - possibly including a nonlinear output term - is defined as generalized residual state-space
neural network (GR-SSNN) in \cite{schoukens_improved_2021}. However, even if the structure of model \eqref{eq:ssnn_schoukens} makes it suitable for system identification/simulation, it does not necessarily ease the design of efficient control laws as the function $f_{n}$ is still nonlinear. 

In recent years, there has been a considerable increase in research related to robustness or stability analysis of neural models. Incremental quadratic constraints are now widely used to analyze \textit{a posteriori} the properties of the neural networks \cite{fazlyab2020safety,hashemi2021certifying} or to train robust models \cite{pauli2022neuralSDP}. These properties are generally written as linear matrix inequality (LMI) constraint problems and included into the 
learning scheme in the form of a barrier term using the logdet matrix function \cite{pauli2022neuralSDP,yin_stability_2021}. This solution raises challenges during the optimization process, the logdet function being undefined outside the LMI feasibility set. One proposed strategy to address this challenge is to implement a backtracking line search, as in \cite{revay2020convex}. However, due to the non-convex nature of the optimization profile, this step may either not converge or induce conservatism.

Inspired by previous works on approximate linearization \cite{hauser1992nonlinear} and data driven linearization \cite{gadginmath2022data}, this paper presents a different strategy by imposing a specific structure for the neural network such that the design of a feedback-linearizing control law trivially follows the learning stage. This facilitates the stability analysis and control design for nonlinear systems.

\subsection{Problem Statement and Contributions}

A way to tackle the issue of efficient design of controller for model \eqref{eq:ssnn_schoukens} is to consider that the term based on the nonlinear function $f_{n}(x(t),u(t))$ is modest compared with the linear part and to consider this term as a bounded disturbance. Thus, the vast literature on linear systems comes at hand to design an adequate controller to cope with these uncertainties to
make them vanish via the robustness property of the feedback control law. However, there is no reason for $f_{n}(x(t),u(t))$ to be small after the identification procedure. On the contrary, if one wishes to minimize the output simulation error,  then minimizing $f_{n}$ would be an antagonistic objective. 

An immediate substitute would be to identify a linear time invariant (LTI) model, but this kind of model is generally not representative enough of the underlying system dynamics. Yet, if one manages to render the dynamic behavior of model \eqref{eq:ssnn_schoukens} linear by feedback, then the design of the control law is made simple. Condition for exact feedback linearization are however often complicated to meet in practice \cite{byrnes1989new} and even to check for neural networks architectures \cite{yesildirek_feedback_1995}.

One solution to overcome this issue is to perform approximate feedback linearization, where the objective is to minimize the residual nonlinearity $f_{n}$. This can be achieved by decomposing it into an input nonlinearity associated with an unavoidable modeling error. The control law can then easily deal with the input nonlinearity and the training procedure ensures that the residual nonlinearity is minimized.

This paper introduces a control-oriented identification procedure, namely \textit{approximately feedback-linearizable state-space neural networks}, to identify system \eqref{eq:modelNL} with a given parameterized evolution equation such that the nonlinearity is explained as much as possible by a nonlinearity on the input. This ensures that the model is approximately linearizable by feedback. This work is an extension of previous work \cite{hacheICSC22} where the nonlinear system was assumed to be exactly feedback linearizable.

\subsection{Notations}
The notations used in this article are standard. For any vector or matrices $x \in \mathbb{R}^n$ or $M \in \mathbb{R}^{n \times m}$, $x^T$ and $M^T$ are their transpose and for $M \in \mathbb{R}^{n \times n}$, $M = M^T > 0$ defines a symmetric positive definite matrix. We use the standard Euclidian norm, $\forall x \in \mathbb{R}^n$, $\|x\| = \sqrt{x^Tx} \in \mathbb{R}$.  Finally, a function $f:\mathbb{R}_{\geq 0} \rightarrow \mathbb{R}_{\geq 0}  $ is a $\mathcal{K}$-function if it is continuous, strictly increasing and $f(0)=0$. In addition, a $\mathcal{K}_\infty$-function is a $\mathcal{K}$-function with $\lim_{x \rightarrow \infty} f(x) = \infty$. Results are presented for  discrete-time systems and $x_+$ stands for the value of 
 $x$ at next time step, i.e. $x_+ = x(t+1)$. In the remainder of this document, time dependence of time varying vectors will be omitted for simplicity when there is no ambiguity.

\section{Approximately feedback-linearizable neural state-space models}

Without loss of generality, function $f_{n}(x,u)$ can be written as the addition of an input nonlinearity $h_{n}(y)$ and an unavoidable residual nonlinearity $g_{n}(x,u)$ such that it holds:
\begin{equation}
f_{n}(x,u) = B h_{n}(y) +  g_{n}(x,u)
\end{equation}
Next section introduces a class of state-space neural networks where the nonlinearity is explained as much as possible by the input nonlinearity $h_{n}(y)$ while we seek to minimize the residual nonlinearity $g_{n}(x,u)$.

\subsection{Definitions}

\textbf{Proposition 1. (AL-SSNN)} We call \textit{Approximately feedback-Linearizable State-Space Neural Networks} nonlinear models with linear output of the form:
\begin{equation}
    \begin{aligned}
    x_+ &= Ax+B\Big(u+h_{n}(y)\Big)+ g_{n}(x,u) \\
    y_\theta &= Cx
    \end{aligned}
    \label{eq:ssnn_flin}
\end{equation}
 with parameters  $\theta= \{A,B,C, h_{n}, g_{n} \}$ and where functions $h_{n}$ and $g_{n}$ are neural networks.

~ \hfill $\blacksquare$

\begin{figure}
    \centering   
    \includegraphics[width=0.6\textwidth]{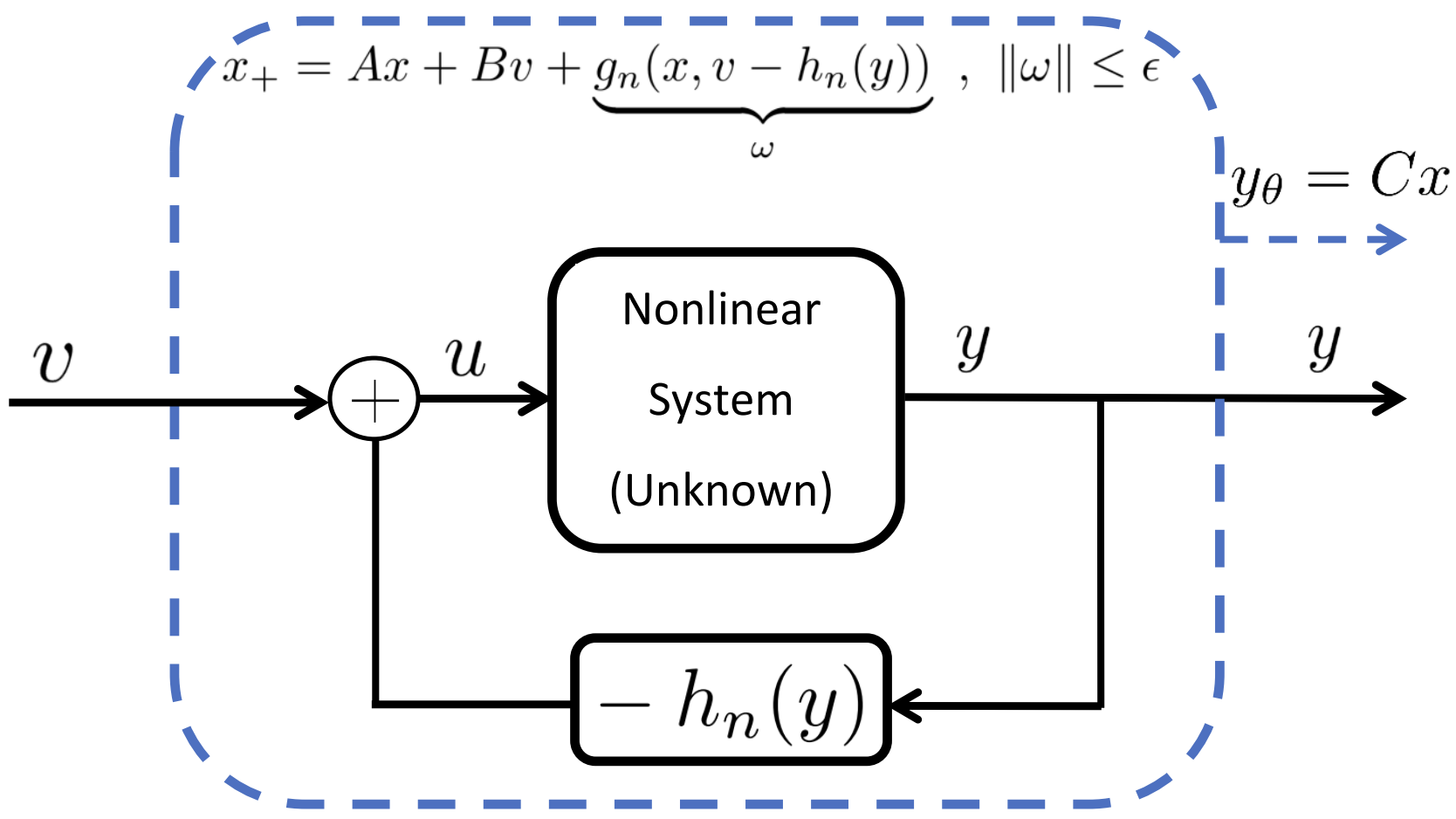}
    \caption{Control architecture for 
 approximate feedback-linearization.}
    \label{fig:schema_rf_gen}
\end{figure}

The interest of this class of model comes from closed-loop control design, as the approximately-linearizing control input trivially writes:
\begin{equation}
    u = v-h_{n}(y)
    \label{eq:quasi_lin_control}
\end{equation}
which yields the corresponding closed-loop model:
\begin{equation}
     \begin{aligned}
    x_+ &=  Ax+Bv +\omega \\
    y_\theta &= Cx
    \end{aligned}
    \label{eq:ssnn_flin_bf}
\end{equation}
which is a LTI model with the closed-loop input $v$ and a bounded perturbation $\omega =  g_{n}(x,v-h_n(y))$, as illustrated on figure \ref{fig:schema_rf_gen}. Control input \eqref{eq:quasi_lin_control} is an output-feedback and therefore does not require the design of a state-observer for the initial nonlinear model.

To ensure that the AL-SSNN model is suitable for control design, the model parameters $\theta$ are obtained through the minimization of the mean-squared error between data measurements and model output, plus the minimization of the residual nonlinearity $g_{n}$:
\begin{equation}
\begin{aligned}
    \hat{\theta} = & \arg \min J_N(\theta) \\
    J_N(\theta) = & \frac{1}{N} \sum_{k=1}^N \Big( \big((y(k)-y_\theta(k)\big)^2   + \gamma g_{n}\big(x(k),u(k)\big)^2 \Big )
   \end{aligned}
    \label{eq:cost_min_res}
\end{equation}
where $y$ is the real system output, $y_\theta$ is the output of model \eqref{eq:ssnn_flin} given parameters $\theta$ and $N$ is the number of samples used for training. Finally, $\gamma$ is a positive scalar that weights the minimization of the nonlinearity compared to the precision of the resulting model.

\section{Identification procedure}
\label{sec:identif}

This section details the implementation of the training procedure of the AL-SSNN model \eqref{eq:ssnn_flin}.

\subsection{Network structure}

Functions $h_{n}(y)$ and $g_{n}(x,u)$ are both represented by a one-hidden-layer feed-forward neural network with nonlinear activation function with respectively $n_h$ and $n_g$ hidden neurons, and a linear output layer:
\begin{equation}
    \begin{aligned}
    h_{n}(y) &= W_h \sigma_h(W_{y} y + b_{y}) + b_h \\
    g_{n}(x,u) &= W_g \sigma_g(W_{x} x + W_{u} u + b_{x}) + b_g 
    \end{aligned}
\end{equation} 
where $\sigma_h : \mathbb{R}^{n_h} \rightarrow \mathbb{R}^{n_h}$ and $\sigma_g : \mathbb{R}^{n_g} \rightarrow \mathbb{R}^{n_g}$ are element-wise nonlinear activation functions that can be set independently but in the following we use hyperbolic tangent as activation function.

The approximately-linearizing control input \eqref{eq:quasi_lin_control} is applied around an equilibrium point $x_e$ induced by $u_e$ and, arbitrarily close to this equilibrium point, the residual nonlinearity $g_n$ vanishes. Its biases terms $b_x$ and $b_g$ are therefore computed such that $g_n(x_e,u_e) = 0$. 

\subsection{Software Implementation}

The cost function \eqref{eq:cost_min_res} is nonlinear and not convex but its gradient can be computed efficiently, which makes the choice of gradient descent algorithm appropriate. The model is trained using algorithm implemented within the Matlab Deep Learning toolbox \cite{demuth1992neural} and Levenberg-Marquardt algorithm \cite{more1978levenberg}. The training configurations for each example is detailed in table \ref{tab:config_training}, the identification method presented here involves only few parameters to tune and does not require the use of deep networks with large number of hidden neurons.

The control-oriented architecture eventually slows down the identification procedure, as illustrated in table \ref{tab:config_training}. Training and simulations tests were carried out on a laptop with an Intel(R) Core(TM) i7 CPU at $1.80$ GHz x 4 with $32$GB of system memory. 

\subsection{Initialization}

 Inspired from \cite{schoukens_improved_2021}, the training procedure is iterative and starts by linear initialization, i.e. the AL-SSNN model \eqref{eq:ssnn_flin} is initialized as:
\begin{equation}
    \begin{aligned}
    x_+ &= A_0x+B_0u \\
    y_\theta &= C_0x 
    \end{aligned}
    \label{eq:lin_init}
\end{equation}
where matrices $A_0$, $B_0$ and $C_0$ are obtained using a linear identification algorithms, namely the subspace method implemented within the Matlab function {\tt ssest} together with {\tt n4sid} algorithm \cite{van1994n4sid}. In other words, functions $h_{n}$ and $g_{n}$ are initialized as $h_{n}(y) = 0$ and $g_n(x,u) = 0$. 

\subsection{Implementation of Cost Function}

Our objective is not only to minimize the nonlinear term $g_n(x,u)$, but rather to minimize the ratio between this nonlinearity and the linear part so that the nonlinear term is negligible in regards to the linear one. In other words, we aim at minimizing  $\frac{\|g_{n}(x,u)\|}{\|Ax+Bu\|}$. However, the computation of the gradient is computationally expensive and may lead to poor training performances.

We propose a different approach to overcome this issue. The minimization of $\|g_{n}(x,u)\|$ is added in the objective function as in \eqref{eq:cost_min_res} and to ensure that $\|Ax+Bu\|$ is not minimized jointly, we impose that the output matrix remains fixed during training, i.e. $C = C_0$. Therefore, the optimization solver cannot scale down $\|x\|$ and scale up matrix $C$ accordingly, otherwise the ratio $\frac{\|g_{n}(x,u)\|}{\|Ax+Bu\|}$ would not be minimized.

\section{Illustrations of training results}

The effectiveness of the proposed identification procedure is illustrated thanks to a popular benchmarks for system identification, a Wiener-Hammerstein process. 

\begin{figure}
    \centering
    \includegraphics[width=0.8\textwidth]{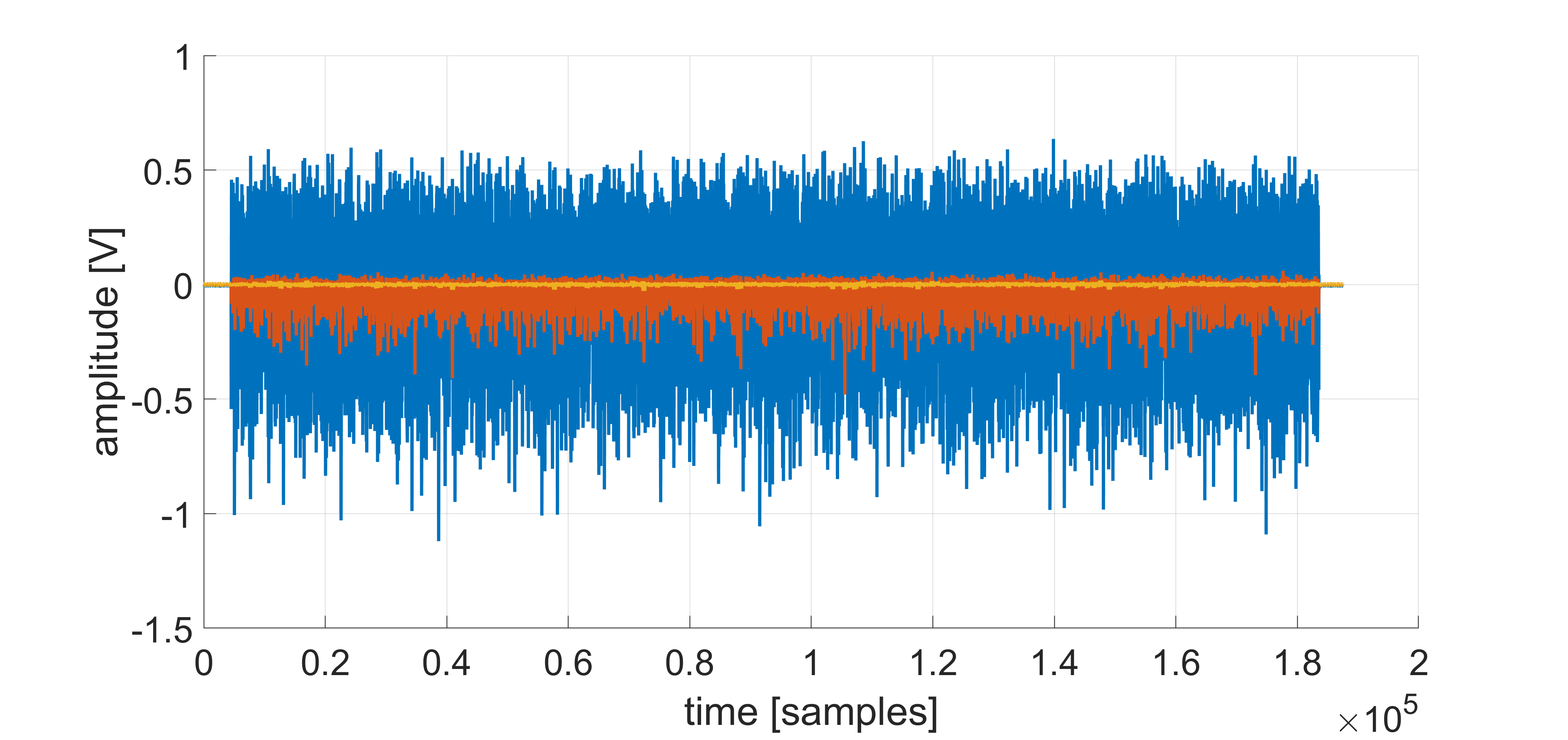}
    \caption{Training results for the Wiener-Hammerstein system. Blue: real output measurements, Red: error between real output and LTI model, Yellow: error between real output and AL-SSNN model \eqref{eq:ssnn_flin}.}
    \label{fig:res_wh}
\end{figure}

\begin{table}
    \centering
    \caption{Wiener-Hammerstein training results. Comparison of root-mean-squared error (RMSE) of the proposed model and two models from literature : generalized ssnn \eqref{eq:ssnn_schoukens} and LTI .}
    \begin{tabular}{c|ccc}
        RMSE & \begin{tabular}{c} AL-SSNN \\ (This work, Eq. \eqref{eq:ssnn_flin}) \end{tabular} &  \begin{tabular}{c} GR-SSNN \\ (Eq. \eqref{eq:ssnn_schoukens}) \end{tabular} & LTI\\ 
        \hline 
        Training & $5.3 \times 10^{-4}$ & $3.6 \times 10^{-4}$ &  $2.7 \times 10^{-2}$ \\
        Test & $5.5 \times 10^{-4}$ & $3.7 \times 10^{-4}$ & $2.8 \times 10^{-2}$
    \end{tabular}
    \label{tab:res_WH}
\end{table}

The Wiener-Hammerstein system is a block-oriented structure composed of a static nonlinearity between two LTI blocks. The underlying differential equations are considered as unknown, we only assume that the state vector is in $\mathbb{R}^6$, following system description \cite{schoukens2009wiener}. 

We seek to identify a AL-SSNN model \eqref{eq:ssnn_flin} from time domain input-output data measurements consisting in $188~000$ points available online\footnote{https://sites.google.com/view/nonlinear-benchmark/benchmarks/wiener-hammerstein}. The data set is decomposed into training (first $100~000$ points) and test (last $88~000$ points).

We compare the identification capabilities of the presented AL-SSNN model \eqref{eq:ssnn_flin} with the one of the generalized nonlinear model \eqref{eq:ssnn_schoukens} where the nonlinear function $f_{n}$ is unconstrained. Results are illustrated on figure \ref{fig:res_wh} that shows the error between the simulated output of the proposed model \eqref{eq:ssnn_flin} and the real output data. In addition, table \ref{tab:res_WH} gathers the root-mean-squared errors (RMSE) between real output measurements and our proposed model. We provide the RMS errors between the data and a LTI model and the generalized nonlinear model \eqref{eq:ssnn_schoukens} for comparison. 

The results indicate that the proposed identification method performs better than standard LTI model and reaches same order of magnitude compared to state-of-the-art nonlinear identification. Our approach performs equally well when $\gamma \rightarrow 0$, it is then a compromise to be found between the models precision and ease of use for control synthesis, keeping in mind that the inaccuracies of the model can be tackled by a closed loop control scheme. A detailed analysis of the influence of parameter $\gamma$ is given in next section in table \ref{tab:diff_gamma_pp}.

\subsubsection{Prey-Predator system}

We consider a discretized version of a three species prey-predator system of the form:
\begin{equation}
    \begin{aligned}
    &\frac{d{x}_1}{dt} = a_1 x_1 - b_1 x_1 x_2 - c_1 x_1 x_3 + d_1 u_1 ^2\\
    &\frac{d{x}_2}{dt} =  a_2 x_2 - b_2 x_1 x_2 - c_2 x_1 x_3 +d_2 u_2^2\\
    &\frac{d{x}_3}{dt} =  -e x_3 + f x_1 x_3 + g x_2 x_3 \\    
    \end{aligned}
\end{equation}
for which the output is the predator $x_3$. The two prey species $x_1$ and $x_2$ are actuated with $u_1^2$ and $u_2^2$. The parameters $a_i$, $b_i$, $c_i$, $d_i$, $e$, $f$ and $g$ represents the growth/death rates, the effect
of predation on the prey population, and the growth of
predators based on the size of the prey population. Inspired by \cite{brunton2016sparse}, for the data generation, we force the system sinusoidally with $u_i(t) = A_i sin(t+\phi_i) + A_i sin(t/10+\phi_i)$ and generate $50~000$ data points from which we use the first half for training and second half for validation.

\begin{figure}
    \centering
    \includegraphics[width=0.8\textwidth]{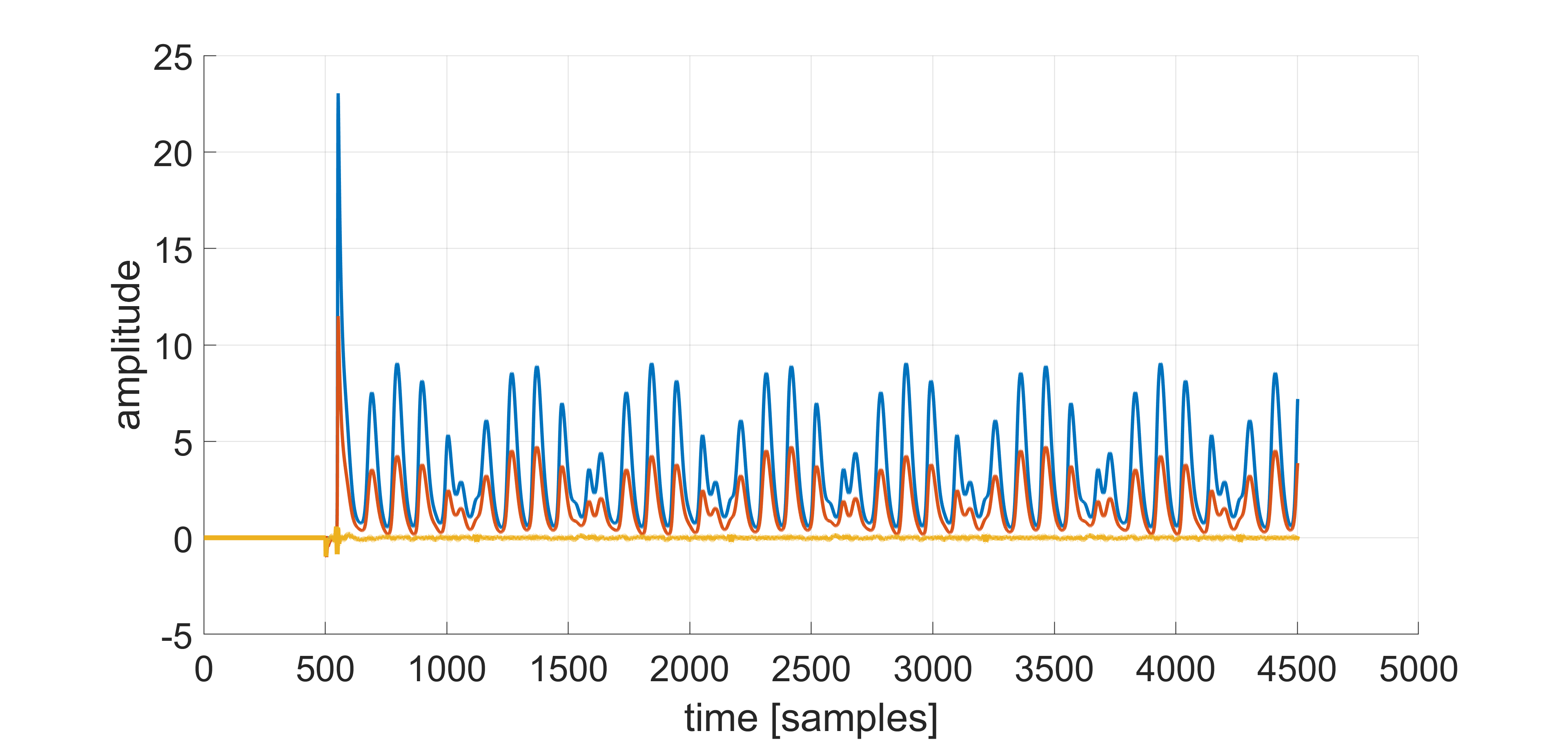}
    \caption{Training results for the prey-predator system.  Blue: output measurements, Red: error between real output and LTI model, Yellow: error between real output and AL-SSNN model  \eqref{eq:ssnn_flin}.}
    \label{fig:res_pp3}
\end{figure}

The identification results are illustrated on figure \ref{fig:res_pp3} and gathered in table \ref{tab:res_pp3}. It shows that the LTI is not able to represent faithfully the underlying dynamics but the identification method proposed above (AL-SSNN, equation \eqref{eq:ssnn_flin}) achieves the same order of magnitude as the generalized nonlinear model used for comparison. In addition, the training procedure ensures that the residual nonlinearity norm $\|g_{n}(x,u)\|$ is minimized. If successful, one should get after training $\|g_{n}(x,u) \| \ll \|Ax+Bu \| $. This is confirmed by results in table \ref{tab:sum_bo} that presents the mean and maximum values of the ratio between the nonlinear and linear terms of equation \eqref{eq:ssnn_flin}. Results show an average ratio of $1.9 \times 10^{-4}$ for the Wiener-Hammerstein example and $2.5 \times 10^{-2}$ for the prey-predator system while it was only of $0.20$ and $0.75$ for the generalized nonlinear model \eqref{eq:ssnn_schoukens}. The proposed approach therefore divide by at least one order of magnitude the norm of the nonlinear residual.

\begin{table}
    \centering
    \caption{Prey-predator training results. Comparison of root-mean-squared error (RMSE) of the proposed model and two models from literature : generalized ssnn \eqref{eq:ssnn_schoukens} and LTI.}
    \begin{tabular}{c|ccc}
        RMSE & \begin{tabular}{c} This work \\ (model \eqref{eq:ssnn_flin}) \end{tabular} &  \begin{tabular}{c} GR-SSNN \\ (model \eqref{eq:ssnn_schoukens}) \end{tabular} & LTI\\ \hline 
        Train & $9.5 \times 10^{-2}$ & $6.7 \times 10^{-2}$ & $1.9 $ \\
        Test & $9.3 \times 10^{-2}$ & $6.5 \times 10^{-2}$ &  $1.9 $
    \end{tabular}
    \label{tab:res_pp3}
\end{table}

\begin{table*}[t]
    \centering
    \caption{Summary of open-loop simulation results for both examples Wiener-Hammerstein (W.H., top) and prey-predator systems (P.P., bottom). Each row depicts the results for the work presented in the paper, AL-SSNN model \eqref{eq:ssnn_flin}, and the generalized version, GR-SSNN \eqref{eq:ssnn_schoukens}, for comparison.}
    \begin{tabular}{c|cc|cc|cc}
          & \multicolumn{2}{c|}{$\frac{\|f_{n}(x,u)\|}{\|Ax+Bu\|}$} 
 & \multicolumn{2}{c}{$\frac{\|g_{n}(x,u)\|}{\|Ax+Bu\|}$} & \multicolumn{2}{c}{$\frac{\|h_{n}(y)\|}{\|Ax+Bu\|}$} \\
         &  $\max $ & ${\tt mean} $ & $\max $ & ${\tt mean} $  & $\max $  & ${\tt mean} $ \\ \hline  
         \begin{tabular}{c}
             W.H., AL-SSNN  \\
             (this work, Eq. \eqref{eq:ssnn_flin}) 
        \end{tabular} & $-$ & $-$ &  $3.6 \times 10^{-3} $ & $1.9 \times 10^{-4}$ & $1.2$ & $5.6 \times 10^{-2}$  \\
        & & & & & &  \\
        W.H., GR-SSNN \eqref{eq:ssnn_schoukens} & $0.45 $ &  $0.20$ &  $-$ & $-$ &  $-$&  $-$  \\  \hline 
                \begin{tabular}{c}
             P.P., AL-SSNN  \\
             (this work, Eq. \eqref{eq:ssnn_flin}) 
        \end{tabular} & $-$ & $-$ & $0.85 $ & $2.5 \times 10^{-2}$ & $25.16$ & $1.00 $  \\
        & & & & & &  \\
        P.P., GR-SSNN \eqref{eq:ssnn_schoukens} & $5.01 $ &  $0.75$  & $-$ & $-$ &  $-$ &  $-$
    \end{tabular}
    \label{tab:sum_bo}
\end{table*}

\section{Closed-loop analysis}
\label{sec:stab}

AL-SSNN model \eqref{eq:ssnn_flin} in closed with $u = v - h_{n}(y)$ is a LTI model with bounded disturbance $\omega$:
\begin{equation}
    \begin{aligned}
    x_+ &= Ax+Bv+\omega \\
    y &= Cx
    \end{aligned}  
    \label{eq:model_bf}
\end{equation}
with $\|\omega\| = \|g_n(x,v-h_n(y)\| \leq \epsilon$. As $\omega$ is bounded and negligible compared to the linear term, the vast theory of linear control theory come at hand to design robust control laws. 

 The training procedure ensures that the \textit{open-loop} residual nonlinearity norm $\|g_n(x,u)\|$ is minimized for all combinaison of $(x,u)$ included in the training set. In the following, it is assumed that minimizing the open-loop norm implies minimizing the \textit{closed-loop} residual nonlinearity $\|g_n(x,v-h_n(y))\|$. This assumption holds if one consider that $v-h_n(y)$ stays in the training set, i.e. the training set is exhaustive enough to capture the system dynamics in its whole. This assumption is verified experimentally in the following sections.

\subsection{Approximate Linearization}

 To estimate how dominant is the linear part of model \eqref{eq:model_bf}, we simulate this model with $v$ being the real data measurements and compare the norm of the linear and nonlinear term of the state equation. If $\|w\|$ is negligible compared to $\|Ax+Bv\|$, then system \eqref{eq:model_bf} is approximately linear, i.e. linear if $\frac{\|w\|}{\|Ax+Bv\|} \rightarrow 0$. Results are gathered in table \ref{tab:sum_bfr} and show that there is on average a ratio of order $10^{-2}$ between the linear and nonlinear term of the Wiener-Hammerstein process. Therefore, the linear part of model \eqref{eq:model_bf} is at least one order of magnitude superior to the residual nonlinearity $\omega$. This ratio can be tuned with the scalar value $\gamma$ in the cost function \eqref{eq:cost_min_res}. If $\gamma \rightarrow 0$, the results are similar with the ones of the generalized model \eqref{eq:ssnn_schoukens}, on the contrary when $\gamma$ increases, then the residual nonlinearity is further minimized, as illustrated by table \ref{tab:diff_gamma_pp}. For training, the value of $\gamma$ is tuned by trial and errors, with the objective of optimizing the compromise between linearization capabilities and precision of the nonlinear model.

 \begin{table}[H]
     \centering
     \caption{Illustration of the impact of the choice of the weighting scalar $\gamma$ in the cost function \eqref{eq:cost_min_res} for the Wiener-Hammerstein process.}
     \begin{tabular}{c|ccc}
     &  $\gamma = 0.01$ & $\gamma = 1$ & $\gamma = 100$ \\ \hline 
         RMS error $(y-y_\theta)$ & $2.8 \times 10^{-4}$ &  $5.5 \times 10^{-4}$ & $7.1 \times 10^{-3}$ \\
         $ {\tt mean} (\frac{\|g_n(x,u)\|}{\|Ax+Bu\|})$ & $5.6 \times 10^{-3}$ & $1.9 \times 10^{-4}$ & $1.9 \times 10^{-5}$ 
     \end{tabular}
     \label{tab:diff_gamma_pp}
 \end{table}

\begin{table*}[t]
    \centering
    \caption{Summary of closed-loop results for both examples Wiener-Hammerstein (W.H., top) and prey-predator (P.P, bottom).}
    \begin{tabular}{c|cc|cc|cc}
          & \multicolumn{2}{c|}{$\|Ax+Bv\|$} 
 & \multicolumn{2}{c|}{$\|g_{n}(x,v-h_{n}(y))\|$} & \multicolumn{2}{c}{$\frac{\|g_{n}(x,v-h_{n}(y))\|}{\|Ax+Bv\|}$} \\
         &  ${\tt max} $ & ${\tt mean} $ & ${\tt max}$ & ${\tt mean} $  & ${\tt max}$  & ${\tt mean} $ \\ \hline 
        W.H. & $8.6$ & $1.9$ & $5.3 \times 10^{-3} $ & $2.7 \times 10^{-4}$  & $4.1 \times 10^{-3}$ & $2.0 \times 10^{-4}$  \\ 
        P.P. & $2.07$ & $0.75$  & $7.8 \times 10^{-2}$ & $1.4 \times 10^{-2}$  & $0.71$ & $1.3 \times 10^{-1}$   
    \end{tabular}
    \label{tab:sum_bfr}
\end{table*}

\subsection{Stability Analysis}

To ensure that the residual nonlinearity do not affect the stability of the closed-loop system, it is taken into account in the stability analysis thanks to input-to-state stability (ISS) property.


 System \eqref{eq:model_bf} is then ISS if it admits a Lyapunov function $V(x)$ such that:
\begin{equation}
    \Delta V(x) < -\Phi(\|x\|) + \Psi(\|\omega\|)
\end{equation}
In the remainder of the document, we use quadratic functions defined as $V(x) = x^TPx$, $\Phi(\|x\|) = \phi x^TP x$ and $\Psi(\|\omega\|) = \psi \omega^T \omega$, where $\phi$ and $\psi$ are two positive scalars.

 Above Lyapunov inequality is then equivalent to:
\begin{equation}
    \begin{aligned}
    & \begin{pmatrix}
        x^T & \omega^T
    \end{pmatrix}
    \begin{pmatrix}
         A^TPA+(\phi-1)P & A^TP \\ PA &  P- \psi I
    \end{pmatrix}
    \begin{pmatrix}
        x \\ \omega
    \end{pmatrix} < 0
    \end{aligned}  
\end{equation}
for which a sufficient condition is the following linear matrix inequality (LMI) constraint problem with decision variables $\phi \in \mathbb{R}_{>0}$, $\psi \in \mathbb{R}_{>0}$, $P >0 \in \mathbb{R}^{n \times n}$ :
\begin{equation}
    \begin{pmatrix}
         A^TPA+(\phi-1)P & A^TP \\ PA &  P- \psi I
    \end{pmatrix} < 0
    \label{eq:lmi_stab}
\end{equation}

It ensures that, for large enough time, the closed-loop state $x$ converges to a hyperball $I$ whose radius depends on $\phi$ and $\psi$:
\begin{equation}
    I = \{~ x \in \mathbb{R}^n,~  \Phi(\|x\|) \leq \Psi(\|\omega\|) ~\}
\end{equation}

Let us now recall that after training,  $\|\omega \| \leq \epsilon $, a specific bound that can be computed on training and validation sets easiliy. Then, it holds $\Psi( \| \omega \| ) = \psi \| \omega \|^2 \leq \psi \epsilon^2$. We finally get the following stability property.

\textbf{Proposition 2.} 

Assuming that:
\begin{itemize}
    \item the AL-SSNN model \eqref{eq:ssnn_flin} approximates system \eqref{eq:modelNL} with $\| g_{n}(x,u)\| \leq \epsilon$
    \item LMI \eqref{eq:lmi_stab} holds
\end{itemize}
then the closed-loop state $x$ of \eqref{eq:model_bf} converges to an invariant manifold defined as:
\begin{equation}
\begin{aligned}
    I &= \{~ x \in \mathbb{R}^n,~ x^TPx \leq \frac{\psi \epsilon^2}{\phi} ~\}
\end{aligned}
\end{equation}
~ \hfill $\blacksquare$

LMI \eqref{eq:lmi_stab} can be computed using semi-definite programming solvers such as YALMIP \cite{yalmip} and the final values of stability  certificates for Wiener-Hammerstein system  are $\phi = 0.1, \psi = 11.$, $\frac{\psi \epsilon^2}{\phi} = 3.1 \times 10^{-4}$ and $\lambda_{min}(P) = 2 \times 10^{-2}$ and $\lambda_{max} = 1.5$.

{\tiny
\begin{table*}[t]
    \centering
    \caption{Summary of training configurations for both examples Wiener-Hammerstein (W.H., top) and prey-predator (P.P., bottom). Each row compares the training configuration for the work presented in the paper, AL-SSNN model, equation \eqref{eq:ssnn_flin} and the generalized version GR-SSNN \eqref{eq:ssnn_schoukens} for comparison.}
    \begin{tabular}{c|ccccccc}
         & \begin{tabular}{c}
              $x \in \mathbb{R}^n$ \\
              $ y \in \mathbb{R}^p$ \\
              $ u \in \mathbb{R}^m$  
         \end{tabular} & $h_{n} \in \mathbb{R}^{n_h}$ & $g_{n} \in \mathbb{R}^{n_g}$ & $f_{n} \in \mathbb{R}^{n_f}$ &  $\gamma$ in \eqref{eq:cost_min_res} & $\#$iter $N_i$ & \begin{tabular}{c}
             Training \\ time T\\
              (min) 
         \end{tabular} \\ \hline   
        \begin{tabular}{c}
             {\scriptsize W.H., AL-SSNN}  \\
             {\scriptsize(this work, Eq. \eqref{eq:ssnn_flin})} 
        \end{tabular}  & \multirow{3}{3em}{$n=6$ \\ $p=1$ \\ $m=1$} & $n_h = 10$ & $n_g = 10$ & $-$ & $\gamma = 1 $ & $N_i = 500$ & $ T = 123$   \\
        & & & & & & & 
        \\
        {\scriptsize W.H., GR-SSNN \eqref{eq:ssnn_schoukens}}  &  & $-$ & $-$ & $n_f = 15$ & $-$ & $N_i = 500$ & $ T = 40.3$   \\ \hline 
        \begin{tabular}{c}
             {\scriptsize P.P., AL-SSNN}  \\
             {\scriptsize (this work, Eq. \eqref{eq:ssnn_flin}) }
        \end{tabular} & \multirow{3}{3em}{$n=3$ \\ $p=1$ \\ $m=2$} & $n_h = 10$ & $n_g = 10$ & $-$ & $\gamma = 2 $ & $N_i= 1000$ & $ T = 36.9 $   \\
        & & & & & & & 
        \\
        {\scriptsize P.P., GR-SSNN \eqref{eq:ssnn_schoukens} }&  & $-$ & $-$ & $n_f = 12$ & $-$ & $N_i = 1000$ & $ T = 19.4$  \\
    \end{tabular}
    \label{tab:config_training}
\end{table*}
}

\section{Discussion}

This contribution is a proof of concept illustrating the possibility of parameterizing and learning a nonlinear model, which opens up the field and its application to practical problems of greater complexity. Results are presented for the linear-output case but can be extended to the general nonlinear case of the form:
\begin{equation}
    \begin{aligned}
    x(t+1) &= Ax(t)+B\Big(u(t)+h_n(y(t)\Big)+g_{n}(x(t),u(t)) \\
    y(t) &= Cx(t) + Du(t)+k_n(x(t),u(t))
    \end{aligned}
\end{equation}
In this case, the optimization problem should be adapted accordingly to include a minimization of the output residual nonlinearity $k_n$:
\begin{equation}
\begin{aligned}
    \hat{\theta} = & \arg \min J_N(\theta) \\
    J_N(\theta) = & \frac{1}{N} \sum_{k=1}^N \Big( (y(k)-y_\theta(k))^2   + \gamma_g g_{n}(x(k),u(k))^2+ \gamma_k k_{n}(x(k),u(k))^2 \Big )
   \end{aligned}
\end{equation}

This would of course require additional computation time and fine tuning of the two scalars $\gamma_f$ and $\gamma_k$ to train the networks. This is the topic of current research.

In addition, only single layer neural network are considered in this work. This is not a restriction as the results directly apply for the general multi-layer case. The two examples of this paper show that single layer neural networks are powerful enough to achieve the desired performances and suggest that single layer networks may be sufficient or even preferable to get models as simple as possible.

\section{Conclusions}

This work proposes a procedure to identify nonlinear state-space models from input-output data. The specificity is that, in addition to faithfully representing the input-output behavior of the plant, its parameterization is devoted to facilitating the design of control laws. From time-domain input-output data measurements, a neural-network is trained to identify a nonlinear discrete-time model that is approximately linearizable by output-feedback. The quasi-linearizing control law derives directly from the identification stage and the residual closed-loop nonlinearity is minimized, which makes it possible to use the classical tools of linear control theory. 

Simulation results on popular and competitive benchmarks available online illustrate the effectiveness and genericity of the approach for robust control. The results are presented here for the discrete-time case but can be adapted to the continuous-time case with few modifications. Our current work focuses on taking into account new constraints in the identification procedure; always with the aim of improving performances and robustness of the control laws designed from data.

%

%
%
\bibliography{templates/bibFile}

\begin{thebibliography}{10}
\providecommand{\url}[1]{#1}
\csname url@samestyle\endcsname
\providecommand{\newblock}{\relax}
\providecommand{\bibinfo}[2]{#2}
\providecommand{\BIBentrySTDinterwordspacing}{\spaceskip=0pt\relax}
\providecommand{\BIBentryALTinterwordstretchfactor}{4}
\providecommand{\BIBentryALTinterwordspacing}{\spaceskip=\fontdimen2\font plus
\BIBentryALTinterwordstretchfactor\fontdimen3\font minus \fontdimen4\font\relax}
\providecommand{\BIBforeignlanguage}[2]{{%
\expandafter\ifx\csname l@#1\endcsname\relax
\typeout{** WARNING: IEEEtran.bst: No hyphenation pattern has been}%
\typeout{** loaded for the language `#1'. Using the pattern for}%
\typeout{** the default language instead.}%
\else
\language=\csname l@#1\endcsname
\fi
#2}}
\providecommand{\BIBdecl}{\relax}
\BIBdecl

\bibitem{alamir2022learning}
M.~Alamir, ``Learning against uncertainty in control engineering,'' \emph{Annual Reviews in Control}, vol.~53, pp. 19--29, 2022.

\bibitem{sznaier2020control}
M.~Sznaier, ``Control oriented learning in the era of big data,'' \emph{IEEE Control Systems Letters}, vol.~5, no.~6, pp. 1855--1867, 2020.

\bibitem{schon2011system}
T.~B. Sch{\"o}n, A.~Wills, and B.~Ninness, ``System identification of nonlinear state-space models,'' \emph{Automatica}, vol.~47, no.~1, pp. 39--49, 2011.

\bibitem{schoukens2019nonlinear}
J.~Schoukens and L.~Ljung, ``Nonlinear system identification: A user-oriented road map,'' \emph{IEEE Control Systems Magazine}, vol.~39, no.~6, pp. 28--99, 2019.

\bibitem{forgione2020model}
M.~Forgione and D.~Piga, ``Model structures and fitting criteria for system identification with neural networks,'' in \emph{2020 IEEE 14th International Conference on Application of Information and Communication Technologies (AICT)}.\hskip 1em plus 0.5em minus 0.4em\relax IEEE, 2020, pp. 1--6.

\bibitem{ghosh2018optimal}
D.~Ghosh, X.~Bombois, J.~Huillery, G.~Scorletti, and G.~Merc{\`e}re, ``Optimal identification experiment design for {LPV} systems using the local approach,'' \emph{Automatica}, vol.~87, pp. 258--266, 2018.

\bibitem{verhoek2022}
C.~Verhoek, G.~I. Beintema, S.~Haesaert, M.~Schoukens, and R.~T{\'o}th, ``Deep-learning-based identification of {LPV} models for nonlinear systems,'' in \emph{2022 IEEE 61st Conference on Decision and Control (CDC)}.\hskip 1em plus 0.5em minus 0.4em\relax IEEE, 2022, pp. 3274--3280.

\bibitem{mauroy2016linear}
A.~Mauroy and J.~Goncalves, ``Linear identification of nonlinear systems: A lifting technique based on the koopman operator,'' in \emph{2016 IEEE 55th Conference on Decision and Control (CDC)}.\hskip 1em plus 0.5em minus 0.4em\relax IEEE, 2016, pp. 6500--6505.

\bibitem{lusch2018deep}
B.~Lusch, J.~N. Kutz, and S.~L. Brunton, ``Deep learning for universal linear embeddings of nonlinear dynamics,'' \emph{Nature communications}, vol.~9, no.~1, pp. 1--10, 2018.

\bibitem{hewing2020learning}
L.~Hewing, K.~P. Wabersich, M.~Menner, and M.~N. Zeilinger, ``Learning-based model predictive control: Toward safe learning in control,'' \emph{Annual Review of Control, Robotics, and Autonomous Systems}, vol.~3, pp. 269--296, 2020.

\bibitem{sjoberg1997estimation}
J.~Sjoberg, ``On estimation of nonlinear black-box models: How to obtain a good initialization,'' in \emph{Neural Networks for Signal Processing VII. Proceedings of the 1997 IEEE Signal Processing Society Workshop}.\hskip 1em plus 0.5em minus 0.4em\relax IEEE, 1997, pp. 72--81.

\bibitem{schoukens_initvieux}
A.~Marconato, J.~Sjöberg, J.~A.~K. Suykens, and J.~Schoukens, ``Improved initialization for nonlinear state-space modeling,'' \emph{IEEE Transactions on Instrumentation and Measurement}, vol.~63, no.~4, pp. 972--980, 2014.

\bibitem{schoukens_improved_2021}
M.~Schoukens, ``Improved initialization of state-space artificial neural networks,'' in \emph{2021 European Control Conference (ECC)}.\hskip 1em plus 0.5em minus 0.4em\relax IEEE, 2021, pp. 1913--1918.

\bibitem{fazlyab2020safety}
M.~Fazlyab, M.~Morari, and G.~J. Pappas, ``Safety verification and robustness analysis of neural networks via quadratic constraints and semidefinite programming,'' \emph{IEEE Transactions on Automatic Control}, vol.~67, no.~1, pp. 1--15, 2020.

\bibitem{hashemi2021certifying}
N.~Hashemi, J.~Ruths, and M.~Fazlyab, ``Certifying incremental quadratic constraints for neural networks via convex optimization,'' in \emph{Learning for Dynamics and Control}.\hskip 1em plus 0.5em minus 0.4em\relax PMLR, 2021, pp. 842--853.

\bibitem{pauli2022neuralSDP}
P.~Pauli, N.~Funcke, D.~Gramlich, M.~A. Msalmi, and F.~Allg{\"o}wer, ``Neural network training under semidefinite constraints,'' in \emph{2022 IEEE 61st Conference on Decision and Control (CDC)}.\hskip 1em plus 0.5em minus 0.4em\relax IEEE, 2022, pp. 2731--2736.

\bibitem{yin_stability_2021}
H.~Yin, P.~Seiler, and M.~Arcak, ``Stability analysis using quadratic constraints for systems with neural network controllers,'' \emph{IEEE Transactions on Automatic Control}, vol.~67, no.~4, pp. 1980--1987, 2022.

\bibitem{revay2020convex}
M.~Revay, R.~Wang, and I.~R. Manchester, ``A convex parameterization of robust recurrent neural networks,'' \emph{IEEE Control Systems Letters}, vol.~5, no.~4, pp. 1363--1368, 2020.

\bibitem{hauser1992nonlinear}
J.~Hauser, S.~Sastry, and P.~Kokotovic, ``Nonlinear control via approximate input-output linearization: The ball and beam example,'' \emph{IEEE transactions on automatic control}, vol.~37, no.~3, pp. 392--398, 1992.

\bibitem{gadginmath2022data}
D.~Gadginmath, V.~Krishnan, and F.~Pasqualetti, ``Data-driven feedback linearization using the koopman generator,'' \emph{arXiv preprint arXiv:2210.05046}, 2022.

\bibitem{byrnes1989new}
C.~I. Byrnes and A.~Isidori, ``New results and examples in nonlinear feedback stabilization,'' \emph{Systems \& Control Letters}, vol.~12, no.~5, pp. 437--442, 1989.

\bibitem{yesildirek_feedback_1995}
A.~Yeşildirek and F.~L. Lewis, ``\BIBforeignlanguage{en}{Feedback linearization using neural networks},'' \emph{\BIBforeignlanguage{en}{Automatica}}, vol.~31, no.~11, pp. 1659--1664, Nov. 1995.

\bibitem{hacheICSC22}
A.~Hache, M.~Thieffry, M.~Yagoubi, and P.~Chevrel, ``Control-oriented neural state-space models for state-feedback linearization and pole placement,'' in \emph{10th International Conference on Systems and Control (ICSC)}.\hskip 1em plus 0.5em minus 0.4em\relax IEEE, 2022, pp. 429--434.

\bibitem{demuth1992neural}
H.~B. Demuth and M.~H. Beale, \emph{Neural network toolbox user's guide}.\hskip 1em plus 0.5em minus 0.4em\relax Mathworks, Incorporated, 1992.

\bibitem{more1978levenberg}
J.~J. Mor{\'e}, ``The {L}evenberg-{M}arquardt algorithm: implementation and theory,'' in \emph{Numerical analysis}.\hskip 1em plus 0.5em minus 0.4em\relax Springer, 1978, pp. 105--116.

\bibitem{van1994n4sid}
P.~Van~Overschee and B.~De~Moor, ``N4sid: Subspace algorithms for the identification of combined deterministic-stochastic systems,'' \emph{Automatica}, vol.~30, no.~1, pp. 75--93, 1994.

\bibitem{schoukens2009wiener}
J.~Schoukens and L.~Ljung, ``Wiener-{H}ammerstein benchmark,'' 2009.

\bibitem{brunton2016sparse}
S.~L. Brunton, J.~L. Proctor, and J.~N. Kutz, ``Sparse identification of nonlinear dynamics with control (sindyc),'' \emph{IFAC-PapersOnLine}, vol.~49, no.~18, pp. 710--715, 2016.

\bibitem{yalmip}
J.~Lofberg, ``Yalmip: A toolbox for modeling and optimization in matlab,'' in \emph{2004 IEEE international conference on robotics and automation (IEEE Cat. No. 04CH37508)}.\hskip 1em plus 0.5em minus 0.4em\relax IEEE, 2004, pp. 284--289.

\end{thebibliography}

\end{document}